\documentstyle[graphicx]{elsart}

\begin{document}

\begin{frontmatter}

\title{Mechanical quality factor of a sapphire fiber at cryogenic temperatures}

\author{
T. Uchiyama\thanksref{tau}}
\author{, T. Tomaru, }
\author{D. Tatsumi\thanksref{tatsu}}
\author{, S. Miyoki, }
\author{M. Ohashi, K. Kuroda}
\address{ICRR, University of Tokyo, 5-1-5 Kashiwanoha, Kashiwa, Chiba 277-8582, Japan}
 
\author{
T. Suzuki, A. Yamamoto, T. Shintomi}
\address{High Energy Accelerator Research Organization (KEK), 1-1 Oho, 
Tsukuba, Ibaraki, 305-0801, Japan} 

\thanks[tau]{Present address: Cryogenics Science Center in KEK, 1-1 Oho, Tsukuba, Ibaraki, 305-0801, Japan}

\thanks[tatsu]{Present address: National Astronomical Observatory (NAO), 2-21-1 Osawa, Mitaka, Tokyo 188-8588, Japan}

\begin{abstract}
A mechanical quality factor of $1.1 \times 10^{7}$ was obtained for the 199\,Hz bending vibrational mode in a monocrystalline sapphire fiber at 6\,K. Consequently, we confirm that pendulum thermal noise of cryogenic mirrors used for gravitational wave detectors can be reduced by the sapphire fiber suspension.  
\end{abstract}


\end{frontmatter}

\section{Introduction}
A cryogenic interferometric gravitational wave detector (Large scale Cryogenic interferometric Gravitational wave Telescope: LCGT) is planned in Japan\cite{lcgt}. The most outstanding characteristic of LCGT is the use of cryogenic techniques to reduce thermal noise. LCGT has fiber suspended pendulum-like mirrors similar to other detectors, such as LIGO\cite{ligo},  VIRGO\cite{virgo}, GEO\cite{geo} and TAMA\cite{tama}. These suspended mirrors behave as test masses for the detection of gravitational waves. The gravitational waves are detected by measuring the difference of displacement between mirrors.

Thermal noise is the thermally excited vibration of the mirrors\cite{thnoise}. Major components of the vibration are the pendulum motion (pendulum thermal noise) and the elastic vibration of the mirror itself (mirror thermal noise). Typically, the pendulum thermal noise is the dominant noise source of the detector in a frequency range from approximately 10\,Hz to 100\,Hz and the mirror thermal noise is dominant from 100\,Hz to 500\,Hz. 

The amplitude of the thermal vibration obeys the equation, 
\begin{equation}
\langle x(\omega)^{2} \rangle \propto T\cdot \phi (\omega),
\label{thnoise:eq} 
\end{equation}
where {\it{T}} is the temperature and $\phi(\omega)$ the dissipation angle. Mechanical quality factor (Q-factor) is defined as the inverse of $\phi(\omega)$ at the resonant frequency $\omega _0$, i. e. $Q = \frac{1}{\phi(\omega _0)}$. Phenomenologically, $\phi(\omega)$ is constant in the frequency range of interest when the dominant source of dissipation is internal friction of the suspension fibers or the mirror substrate. This is known as structure damping\cite{thnoise}. For measurement analysis, we applied the structure damping model and focused attention on the Q-factor. 

According to Eq.\ref{thnoise:eq}, the reduction of the thermal noise is accomplished both by lowering the temperature and by improving the Q-factors of both the pendulum and mirror acoustic modes. Our strategy to reduce the thermal noise includes cooling the whole mirror suspension system to cryogenic temperature (below 30\,K at the mirror), along with the use of low acoustic loss material for the mirror and suspension fiber. We have chosen sapphire for both the mirror and fiber, since it has shown to have a high Q-factor\cite{qsapphire} and high thermal conductivity\cite{sconductivity} at cryogenic temperatures. We have already presented results of the cryogenic mirror suspension system with a sapphire test mass which was cooled down to cryogenic temperatures and exhibited a Q-factor of $10^{8}$\cite{coolingmirror,mirrorq}. These results suggests that the cryogenic mirror suspension system could possibly reduce the mirror thermal noise.

To enhance the sensitivity of the detector, we must also reduce the pendulum thermal noise which is estimated by the Q-factor of the pendulum motion ($Q_{pend}$) with temperature. Since $Q_{pend}$ depends on the dissipation in the suspension fibers, we must construct them of low loss materials. Since direct measurement of $Q_{pend}$ is not an easy task, we measured the frequency dependence of Q-factor of the sapphire fiber itself ($Q_{fiber}$). $Q_{pend}$ can be obtained by applying the result of the dissipation dilution theorem to $Q_{fiber}$\cite{thnoise}.

The measurement was undertaken at cryogenic temperatures (6\,K and 78\,K) with one end of the fiber fixed in a clamp and the other end free. We chose the lowest frequency vibration bending mode of the clamped sapphire fiber, since the deformation of this mode is similar to that of the pendulum motion. Applying the structure damping model, the dissipation in the sapphire fiber due to bending vibration reflects that due to pendulum motion, even though the resonant frequencies are different (the resonance of the pendulum is 1\,Hz, whereas that of the fiber is 200\,Hz to 500\,Hz). 

Room temperature measurement of $Q_{fiber}$ of the sapphire fiber represents $5 \times 10^{3}$ at 200\,Hz\cite{qfiber}. This Q-factor measurement was limited by the thermoelastic effect\cite{zener}. Since the thermoelastic effect of the sapphire fiber is negligibly small in cryogenic temperature, $Q_{fiber}$ is expected to be higher for this measurement. To our knowledge, this paper represents the first cryogenic measurement of sapphire fiber Q-factor.

In this letter, we show that the measured $Q_{fiber}$ of the sapphire fiber at 6\,K was as high as that exhibited at room temperature by a fused silica fiber considered to be the highest Q-factor fiber at room temperature.

\section{Experiment and Results}
\begin{figure}
\begin{center}
\includegraphics[height=8cm]{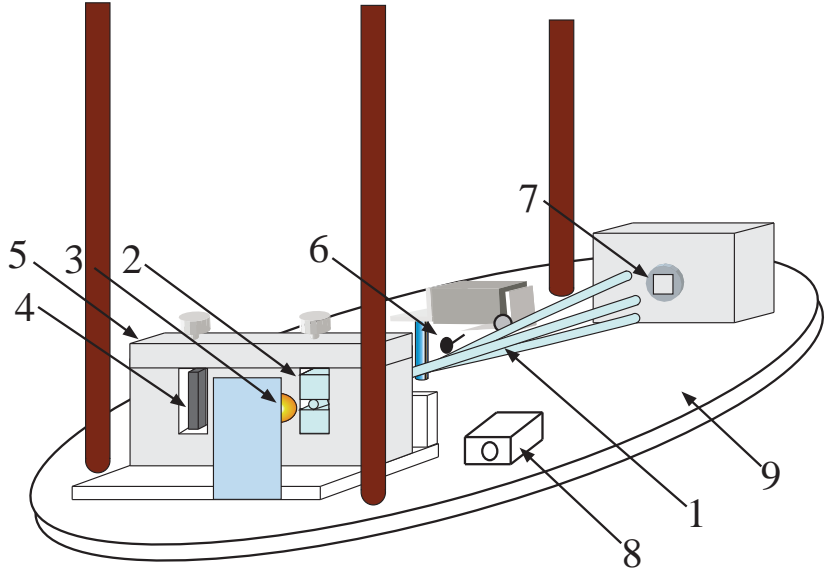}
\end{center}
\caption{\small{Schematic view of set up of the experiment. 1) sapphire fiber, 250\,${\rm \mu m}$ in diameter; 2) sapphire block, $5\,\mathrm{mm} \times 5\,\mathrm{mm} \times 10\,\mathrm{mm}$; 3) infrared light-emitting diode (IR LED); 4) PZT vibrator for continuous excitation; 5) clamp base made of aluminum; 6) hammer for pulse excitation; 7) InGaAs photo diode; 8) copper block for housing a carbon-glass residual (CGR) thermometer; 9) copper table fixed to the top flange of the vacuum chamber with three copper rods. Although this table was expected to be at 4.2\,K, measurements were undertaken at 6\,K due to heat produced by the IR LED. The sapphire fiber was clamped through the sapphire blocks by a M4 aluminum screw with torque of 50\,N$\cdot$cm. The amplitude of the free oscillation was monitored by a shadow sensor that consisted of the IR LED and the InGaAs photo diode. Two methods for excitation were prepared. When searching the resonance, the hammer that provided a large excitation was used. The PZT vibrator was used for exciting a particular mode in the decay time measurement.}}
\label{setup}
\end{figure}

Figure \ref{setup} shows the experimental setup. We used a Saphikon Inc. sapphire fiber\cite{saphikon}, 250\,$\mu$m in diameter, the same as used in previous experiments\cite{coolingmirror,mirrorq}. The whole assembly was housed in a vacuum chamber at $10^{-4}$Pa. The vacuum chamber was immersed in liquid helium in a top-loading cryostat. The whole assembly was cooled by heat conduction through three copper rods from the top flange of the vacuum chamber.

\begin{figure}
\begin{center}
\includegraphics[height= 8cm]{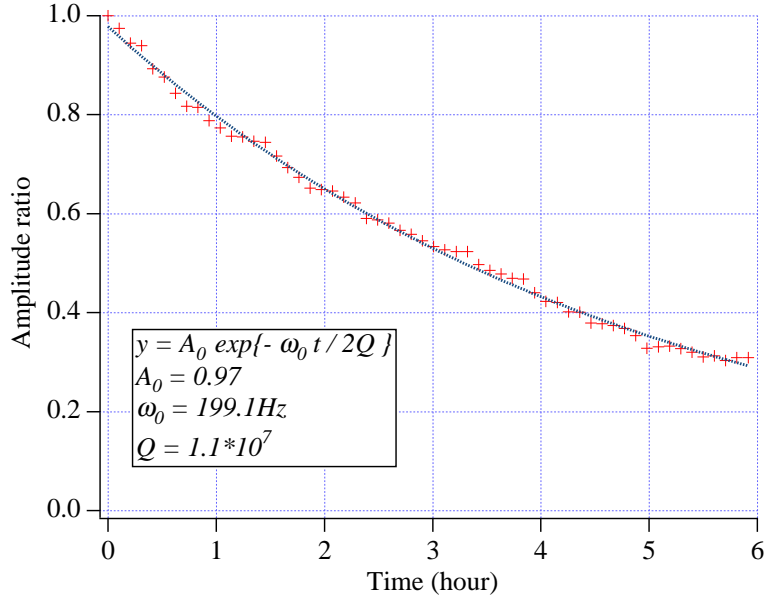}
\end{center}
\caption{\small{The decay of the amplitude of the measurement. Q-factor was calculated by curve fitting to the decay time of free oscillation. }}
\label{example}
\end{figure}

To measure $Q_{fiber}$, the ring-down method was used. A PZT vibrator excited the bending motion of the sapphire fiber. Damping time of the oscillation was measured by a shadow sensor consisting of an infrared light-emitting diode (IR LED) and an InGaAs photo diode. The IR LED was set on the back side of a clamp. The oscillating sapphire fiber cast a shadow on the surface of the photo diode with the size of the shadow changing with the vibration amplitude. Figure \ref{example} shows the amplitude decay in a typical measurement.

We used a pair of the sapphire blocks\cite{kyocera} of 5\,mm in thickness for clamping the sapphire fiber. When the sapphire fiber was clamped with soft metal plates, grooves were formed on the plates and the groove could cause excess loss. The clamp had two slots. One slot was used for clamping the sapphire fiber and the other slot was for setting a PZT vibrator as a continuous exciter. The sapphire fiber was mounted between the sapphire blocks and fastened by a M4 aluminum screw with a torque of 50\,N$\cdot$cm. 

To obtain the $Q_{fiber}$ of other frequency modes, we cut the free end of the clamped sapphire fiber without loosening the clamp and  re-measured $Q_{fiber}$. Thus, the clamping condition remained unchanged for the whole measurement. We found two resonances whose frequencies were separated by about 3\% for each length of the fiber. This frequency split may be due to come from an imperfection circular cross section of the sapphire fiber. In this paper, we distinguish these resonances by their resonant frequencies and refer to them as "Higher resonance" and "Lower resonance". 

\begin{figure}
\begin{center}
\includegraphics[height= 8cm]{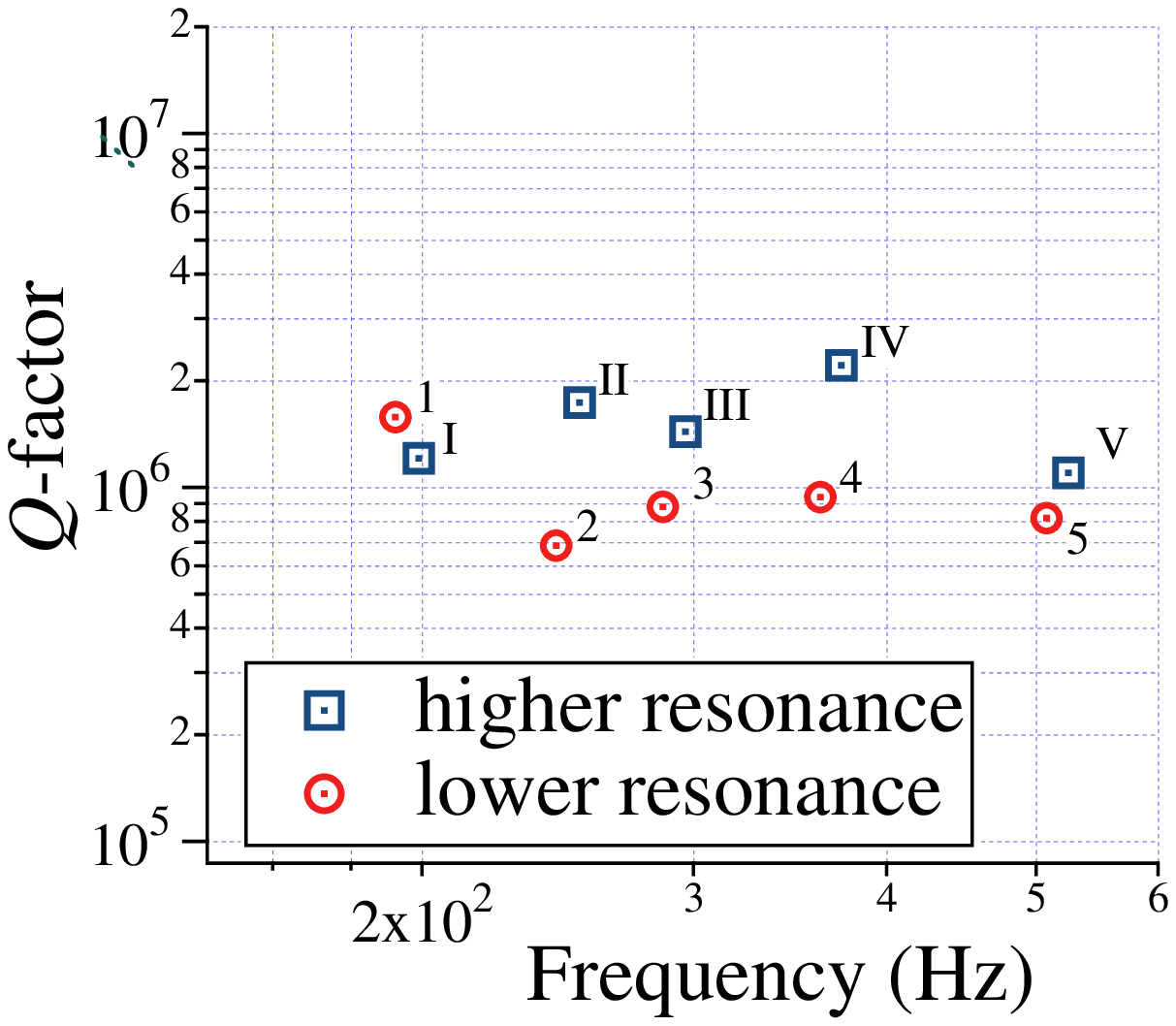}
\end{center}
\caption{\small{Frequency dependence of the Q-factor of the bending mode of the clamped sapphire fiber at 78\,K. A "Higher resonance" and "Lower resonance" pair, indicated by the same number shows a measurement made of under the same fiber length. The numbers correspond to "Number" in Tables \ref{resonantfreq_low:ta} and \ref{resonantfreq_high:ta}.}}
\label{result_n2}
\end{figure}

\begin{figure}
\begin{center}
\includegraphics[height= 8cm]{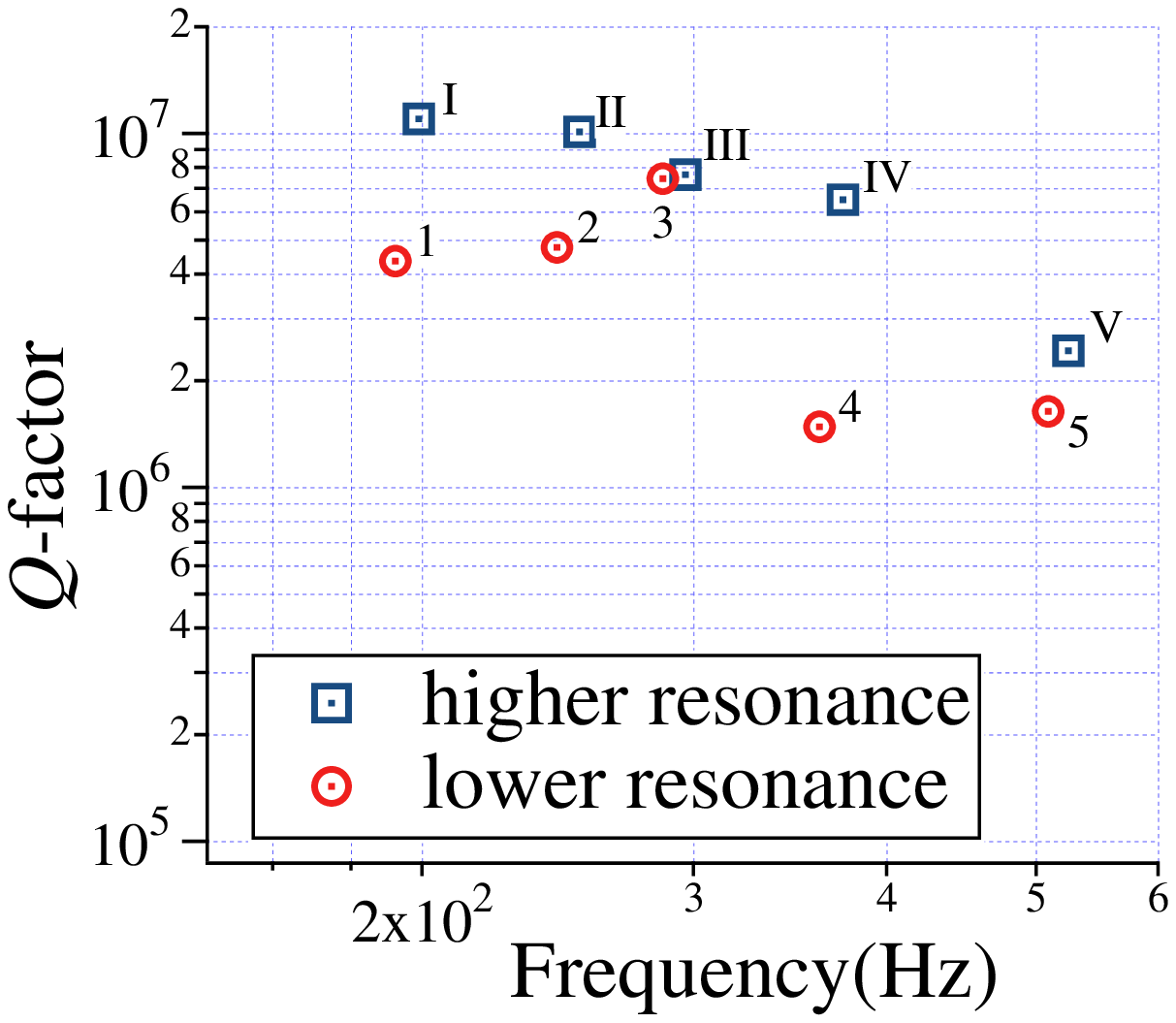}
\end{center}
\caption{\small{Frequency dependence of the Q-factor of the bending mode of the clamped sapphire fiber at 6\,K. A "Higher resonance" and "Lower resonance" pair, indicated by the same number shows a measurement made of under the same fiber length. The numbers correspond to "Number" in Tables \ref{resonantfreq_low:ta} and \ref{resonantfreq_high:ta}.}}
\label{result_he}
\end{figure}

Figure \ref{result_n2} and Figure \ref{result_he} show the frequency dependence of the $Q_{fiber}$ at 78\,K and 6\,K, respectively. The length of the sapphire fiber, the measured resonant frequencies and $Q_{fiber}$ are summarized in Table \ref{resonantfreq_low:ta} and \ref{resonantfreq_high:ta}.  When the sapphire fibers are applied to LCGT as mirror suspension fibers, temperature at clamping points of the sapphire fibers is designed to be lower than 10\,K. Thus, the 6\,K results of $Q_{fiber}$ have more importance to LCGT.

The result of $Q_{fiber}$ at 6\,K shows a frequency dependence. Estimated $Q^{-1}$ due to the thermoelastic effect of the sapphire fiber at 200\,Hz was $5 \times 10^{-16}$, much smaller than the measured value. Although the major dissipation mechanisms have not yet been identified in our measurement, a combination of the sapphire fiber and the clamp using the sapphire blocks has achieved a high $Q_{fiber}$ of more than $10^6$, comparable with $Q_{fiber}$ of fused silica fiber\cite{qfiber}.

For estimating the thermal vibration of the pendulum, we took $Q_{fiber}$ of $7.7 \times 10^{6}$ which is an averaged value of the results at 6\,K at the lowest frequency (192\,Hz and 199\,Hz). Assuming the structure damping model, the value of $Q_{fiber}$ at 199\,Hz can be extrapolated to the frequency region below 100\,Hz where the pendulum thermal noise is a dominant noise source of the detector. $Q_{pend}$ of the cryogenic mirror suspension is obtained through the dissipation dilution theorem\cite{thnoise}. A brief review of the dissipation dilution theorem can be found in an Appendix to this letter. 

\begin{table}
\begin{center}
\caption{\small{Measured resonant frequencies and $Q_{fiber}$ of the lower resonance mode. }}
\label{resonantfreq_low:ta}
\begin{tabular}{|c||c|c|c|c|} \hline
Number & length of the fiber & resonant frequency & $Q_{fiber}$ at 78\,K & $Q_{fiber}$ at 6\,K  \\ \hline
1 & 43.0\,mm & 192\,Hz & 1.6 $\times 10^{6}$ & 4.4 $\times 10^{6}$ \\ \hline
2 & 38.4\,mm & 245\,Hz & 6.8 $\times 10^{5}$ & 4.8 $\times 10^{6}$ \\ \hline
3 & 35.5\,mm & 287\,Hz & 8.8 $\times 10^{5}$ & 7.4 $\times 10^{6}$ \\ \hline
4 & 31.0\,mm & 362\,Hz & 9.4 $\times 10^{5}$ & 1.5 $\times 10^{6}$ \\ \hline
5 & 26.6\,mm & 508\,Hz & 8.2 $\times 10^{5}$ & 1.6 $\times 10^{6}$ \\ \hline
\end{tabular}
\end{center}

\end{table}%

\begin{table}
\begin{center}
\caption{\small{Measured resonant frequencies and $Q_{fiber}$ of the higher resonance mode. }}
\label{resonantfreq_high:ta}
\begin{tabular}{|c||c|c|c|c|} \hline
Number & length of the fiber & resonant frequency & $Q_{fiber}$ at 78\,K & $Q_{fiber}$ at 6\,K \\ \hline
I & 43.0\,mm  & 199\,Hz & 1.2 $\times 10^{6}$ & 1.1 $\times 10^{7}$ \\ \hline
II & 38.4\,mm  & 253\,Hz & 1.7 $\times 10^{6}$ & 1.0 $\times 10^{7}$ \\ \hline
III & 35.5\,mm  & 296\,Hz & 1.4 $\times 10^{6}$ & 7.6 $\times 10^{6}$ \\ \hline
IV & 31.0\,mm  & 374\,Hz & 2.2 $\times 10^{6}$ & 6.5 $\times 10^{6}$ \\ \hline
V & 26.6\,mm  & 525\,Hz & 1.1 $\times 10^{6}$ & 2.4 $\times 10^{6}$ \\ \hline
\end{tabular}
\end{center}

\end{table}%

\section{Discussion}
\label{discussion}

\begin{figure}
\begin{center}
\includegraphics[height= 6cm]{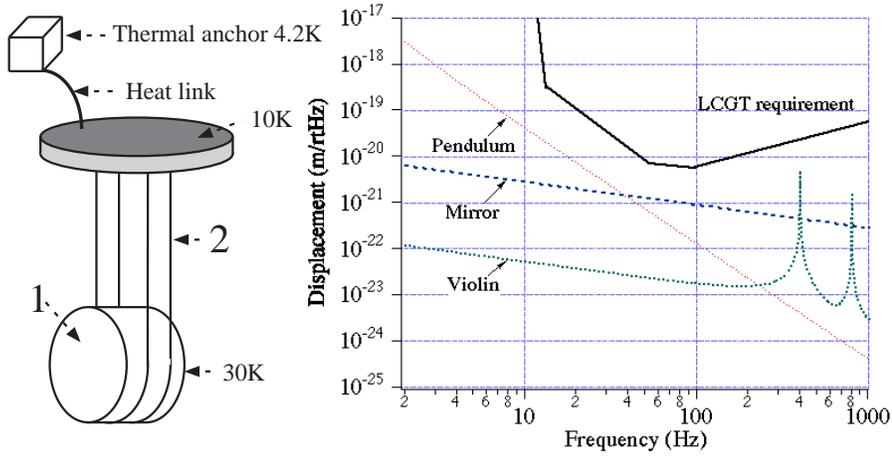}
\end{center}
\caption{\small{Left side: A present conception of the final stage of a cryogenic mirror suspension for LCGT (1) Sapphire mirror (300\,mm in diameter, 180\,mm in length and 50\,kg in mass); (2) Sapphire fiber (1\,mm in diameter and 250\,mm in length, that is 1\,Hz pendulum). Temperature of the mirror and at clamping points of the fibers are assumed to be 30\,K and 10\,K, respectively. Right side: Comparison between estimated noise curves due to the thermal noise and the requirements for LCGT\cite{lcgt}. The $Q_{pend}$ and the Q-factor of the sapphire cylinder uses in this estimation were assumed to be $2.9 \times 10^8$ and $7.2 \times 10^7$\cite{mirrorq}, respectively.}}
\label{estimation}
\end{figure}

Three experiments for the cryogenic mirror suspension system have been done including this measurement and all results of the experiments are promising for realizing a practical cryogenic interferometer. Figure \ref{estimation} shows the present concept of the final stage of a cryogenic mirror suspension for LCGT and a comparison between the estimated thermal noise curves and the requirements of LCGT\cite{lcgt}. The target requirement sensitivity of LCGT is ten times better than LIGO or VIRGO\cite{lcgt}. Figure \ref{estimation} shows that the estimated thermal noise of the LCGT suspension is much lower than the requirement.

The LCGT suspension consists of a sapphire mirror (300\,mm in diameter, 180\,mm in thickness and mass of 50\,kg) that is three times larger one used in Reference \cite{mirrorq}, and two loops of sapphire fibers (1\,mm in diameter and 250\,mm in length of pendulum). Q-factor of the mirror for this estimation used the value of "longitudinal mode (68\,kHz)" in Reference \cite{mirrorq} at 30\,K, (i. e. $7.2 \times 10^7$). In the estimation of the thermal vibration of the mirror, we considered contributions not only from the longitudinal mode but also other elastic modes of the mirror\cite{gillespie}. The resonant frequency of the pendulum was assumed to be 1\,Hz and $Q_{pend}$ of $2.9 \times 10^8$ was obtained as shown in the Appendix.

We assumed that the temperature of the mirror was 30\,K and at the clamping points of the fibers were 10\,K. The temperature of the mirror was limited by temperature dependence of the thermal conductivity of the sapphire fiber which allows a worst-case at 30\,K. The temperature of the mirror was determined by a balance of absorption of laser power in the mirror and the heat conduction of the sapphire fiber. Since the thermal conductivity of sapphire decreases above 30\,K, the heat flux through the fibers does not increase, even if the temperature of the mirror becomes higher than 30\,K. Thus, 30\,K is an upper limit of the temperature of the "cryogenic" mirror.

\section{Conclusion}
We measured the $Q_{fiber}$ of the sapphire fiber in cryogenic temperatures. The highest value of $1.1 \times 10^{7}$ was obtained at 199\,Hz. The measurement showed frequency dependence, however, all measurements of $Q_{fiber}$ at 6\,K achieved over $10^6$ which was as high as $Q_{fiber}$ of fused silica fiber. Since temperature at the clamping points of the suspension fiber in the cryogenic mirror suspension system is considered to be below 10\,K, the pendulum thermal noise of this suspension system is improved from that of a conventional room temperature suspension system.

The cryogenic mirror suspension system may be the lowest thermal noise system for interferometric gravitational wave detectors. These results are encouraging for realizing the LCGT project.

\ack
This study was supported by the Joint Research and Development Program of KEK and by a grant-in-aid prepared from the Ministry of Education, Science, Sports and Culture.

\appendix{Appendix}

\section{Dissipation dilution theorem}
To obtain $Q_{pend}$ for the estimation mentioned in Part \ref{discussion}, the dissipation dilution theorem\cite{thnoise} was applied to $Q_{fiber}$ of the suspension system. Energy dissipation $\Delta E$ due to the pendulum motion occurs in the suspension fiber. In this case, $Q_{pend}$ can be written as 
\begin{equation}
\label{qpend:eq}
Q_{pend} = \frac{E_{el} + E_{gr}}{\Delta E}
\end{equation}
where $E_{el}$ is elastic energy stored in the fiber and $E_{gr}$ is energy due to the gravity force. Generally, $E_{el} \ll E_{gr}$ and $E_{gr}$ is loss-less energy, so $Q_{fiber}$ can be represented by 
\begin{equation}
\label{qbend:eq}
Q_{fiber} = \frac{E_{el}}{\Delta E}.
\end{equation}
From both Eq. (\ref{qpend:eq}) and Eq. (\ref{qbend:eq}), $Q_{pend}$ can be calculated from $Q_{fiber}$ as follows.
\begin{equation}
Q_{pend} \approx Q_{fiber} \times \frac{E_{gr}}{E_{el}},
\end{equation}
\begin{equation}
\eta = \frac{E_{gr}}{E_{el}} = \frac{mg/l}{n\sqrt{TYI}/2l^2},
\end{equation}
where $m$ is mass of the mirror, $g$ is acceleration of the gravity, $l$ is length of the fiber, $n$ is number of the fiber, $T$ is tension applying to the fiber, $Y$ is Young's modulus of the fiber and $I$ is area moment of the inertia of the fiber, respectively. The area moment of the inertia is given by diameter of the fiber $"d"$ as $I = \frac{\pi d^4}{64}$. These parameters are summarized in Table \ref{parameter}.

Under these conditions, the dissipation dilution ratio $\eta ^{-1}$ becomes $2.7 \times 10^{-2}$. Thus ${Q_{pend}}$ of $2.9 \times 10^8$ was estimated for the LCGT suspension. 

\begin{table}[htdp]
\caption{\small{Summary of the parameters for the estimation of $Q_{pend}$.}}
\begin{center}
\begin{tabular}{|c|c|c|} \hline
Objects & symbol & value \\ \hline
mass of the mirror & $m$ & 50\,[kg]\\
gravity acceleration & $g$ & 9.8\,$\mathrm{[m/s^2]}$ \\
length of the pendulum & $l$ & 0.25\,[m] \\
number of the fiber & $n$ & 4(2 loops) \\
diameter of the fiber & $d$ & 1\,[mm] \\
Young's modulus & $Y$ & $4.7 \times 10^{11}\mathrm{[N/m^2]}$ \\ \hline
dissipation dilution ratio & $\eta ^{-1}$ & $2.7 \times 10^{-2}$ \\ 
Q-factor of the sapphire fiber & $Q_{fiber}$ & $7.7 \times 10^6$ \\
Q-factor of the pendulum & $Q_{pend}$ & $2.9 \times 10^8$ \\ \hline
\end{tabular}
\end{center}
\label{parameter}
\end{table}%

\end{document}